\documentclass[12pt,column]{article}

\usepackage{scicite}
\usepackage{subfig}
\usepackage{graphicx}
\usepackage[countmax]{subfloat}
\usepackage{times}
\usepackage{color}
\usepackage{amsmath}

\newenvironment{sciabstract}{%
\begin{quote} \bf}
{\end{quote}}

\title{Topological resilience in non-normal networked systems}

\author
{Malbor Asllani,$^{1\ast}$ Timoteo Carletti,$^{1}$\\
\\
\normalsize{$^{1}$Department of Mathematics \& naXys, Namur Institute for Complex Systems, University of Namur,}\\
\normalsize{rempart de la Vierge 8, B 5000 Namur, Belgium}\\
\\
\normalsize{$^\ast$To whom correspondence should be addressed; E-mail:  malbor.asllani@unamur.be.}
}

\date{}

\begin{document} 


\baselineskip24pt


\maketitle


\begin{sciabstract}
The network of interactions in complex systems, strongly influences their resilience, the system capability to resist to external perturbations or structural damages and to promptly recover thereafter. The phenomenon manifests itself in different domains, e.g. cascade failures in computer networks or parasitic species invasion in ecosystems. Understanding the networks topological features that affect the resilience phenomenon remains a challenging goal of the design of robust complex systems. We prove that the non-normality character of the network of interactions amplifies the response of the system to exogenous disturbances and can drastically change the global dynamics. We provide an illustrative application to ecology by proposing a mechanism to mute the Allee effect and eventually a new theory of patterns formation involving a single diffusing species.
\end{sciabstract}


\section*{Introduction}

The ecological resilience \cite{resilienceeco} is the ability of ecosystems to respond to challenges such as fires, windstorms, deforestation, flooding or the presence of invasive species, and their aptitude to return close to the initial state. The way rapid climate changes are affecting the natural habitats \cite{climate} and how the increasing human activity has been responsible for environmental disasters \cite{socioeco} are indeed the focus of recent studies. On the other hand, resilience is encountered also in human-made systems such as power grids or communications systems where a failure of a component of a system of interconnected elements can trigger a cascade of failures of successive components \cite{cascade}. In this case, the response of the system is directly correlated to the structural changes in the networked support where the dynamics occurs \cite{barabasi}. Efforts have been made to understand the role that the topology of the interactions has on the resilience \cite{havlin} in order to optimize the design that enhances the robustness of such systems and reduce their vulnerability \cite{asha}.  

Here we show that the resilience of dynamical systems evolving on a complex network is highly determined by the degree of non-normality that characterizes its network of interactions. This technical definition~\cite{spectra}, based on the non-existence of a suitable orthogonal basis, will be proved in the following to determine an unexpected system response to small disturbances. We hereby anticipate that this abnormal behavior follows a transient amplification process in the initial linear regime which, if sufficiently large, leads subsequently the system to another state possibly characterized by a spatially inhomogeneous pattern {and potentially far from the initial one, reducing thus the system resilience}. The non-normality effect has been previously studied in different domains such as hydrodynamic stability \cite{pseudospectra}, non-Hermitian quantum-mechanics \cite{hatano} or ecology \cite{ecoresilience}. In particular in population dynamics, it was shown that particular conditions among the inter-species interactions, allow the formation of spatial patterns \cite{neubertmurray,tommaso}. {Let us stress that, differently form the latter cases where the assumption of non-normality relies on strict equality conditions among the parameters of the model, in our approach is the network structure that encodes such feature, allowing thus to potentially consider applications to systems where the geometry of the spatial interactions plays a relevant role.} So far, the only case when the latter relevant peculiarity has been taken in account in relation to complex networks, is in the study of excitatory-inhibitory neural circuits \cite{neurons}. We thus aim to bring to the fore a general framework, where the impact of the non-normality assumption on the system resilience, will emerge straightforwardly. A major example hereby developed that illustrates the importance of this powerful mechanism, is the deviance of the Allee effect \cite{allebook}. The principle according to which initial low densities of a given species may critically endanger its survivability as firstly observed by W.C. Allee{, who remarked that goldfishes grow more rapidly when there are more individuals in a given reservoir \cite{allee}; observation that allowed him to draw the conclusion that segregation and cooperation can improve the survival rate.} We show that the transient growth, induced by the non-normal network of interactions, may reverse this property. Furthermore, because of its generality, we suggest that this new framework could be an alternative pathway to the emergence of spatially self-organized heterogeneous patterns in a Turing-like fashion \cite{murray}. Turing's theory states that a minimal system of activator-inhibitor individuals can generate complex patterns, the celebrated Turing patterns \cite{turing}, following a diffusion-induced instability. Indeed, it is possible to further simplify the pattern forming condition to a single inhibitor species allowed to freely diffuse on a non-normal network whose topological properties will cause a transient instability, strong enough for the non-homogeneous patterns to emerge.

\section*{Non-normal dynamics in networked systems}

Complex interactions in systems, usually constituted by a large number of components, can often be encapsulated in a graph representation through the adjacency matrix $\textbf{A}$ whose entries $A_{ij}=1$ if the node $j$ is directly connected to node $i$ and zero otherwise~\cite{newman}. If also the node $i$ is directly connected to node $j$ determines if the network is directed or not, a condition that can have a strong impact on the system dynamics. 

Let us consider the following generic dynamical system made of $M$ non-linearly coupled components:
\begin{equation}
\frac{dx_i}{dt}=f(x_i) + \sum_{j=1}^M A_{ij}g(x_i,x_j)\, ,
\label{eq:general}
\end{equation} 
where $\textbf{x}=(x_1,x_2,\dots,x_M)$ is the vector of the system states and $f(\cdot)$, respectively $g(\cdot,\cdot)$, is a non-linear function defining the local dynamics occurring on each node $i$, respectively the interaction dynamics triggered by the network topology encoded into the adjacency matrix $\textbf{A}$. Obviously, the evolution of the states of the system will directly depend on the structure of network. Despite the non-linear nature of the system, one can often obtain a good description by linearizing close to an equilibrium solution of Eq.~(\ref{eq:general}); for a sake of completeness, let us assume a diffusive-like {non-linear} coupling, namely $g$ depends on $x_i-x_j$ and consider thus a linear model of a diffusion process on a network with sinks located at each node, namely $\dot{\textbf{x}}=-a\textbf{x} + D\textbf{Lx}$, where $a$ is the decaying rate (assumed for simplicity to be the same for all nodes) and $\textbf{L}$ the discrete Laplacian matrix, $\textbf{L}=\textbf{A}-\textbf{k}^{out}$, where $\textbf{k}^{out}$ is the diagonal matrix of the connectivities, namely the number of outgoing edges from each node.
If the matrix $\textbf{A}$ is non-Hermitian, and so $\textbf{L}$ does, then a basis formed by linearly independent eigenvectors  may not exist; we are hence dealing with a non-normal matrix~\cite{spectra} and the spectrum of the eigenvalues fails to capture the linear dynamics behavior occurring on such network which deviates from the trivial exponential decay. We define such behavior the \textit{topological resilience} and throughout this paper we will refer to such networks, as \textit{non-normal} networks.

A simple but prototypical example for studying the resilience response of the system in non-normality conditions, is shown in Fig. 1 where we represent two networks build on a unidirectional ring. The first network has a full rotational symmetry, each vertex has one incoming link and an outgoing one. When the system is perturbed around its unique and stable state $\textbf{x}^*=(0,\dots,0)$, independently of the initial conditions or of the size of the network, the system converges asymptotically to zero (see top panel of Fig. 1). However, if we remove just one link (as in the second network) and the diffusion is stronger then the decay, $D/a >>1$, we observe that the unique node with just a single incoming link, will exhibit a transient growth before it turns again toward the zero state (see bottom panel of Fig. 1). We can intuitively explain such behavior, by observing that if the individuals diffuse much faster than they are removed from the system due to the sinks, they first will accumulate in the node from where they cannot exit anymore and thereafter, on a longer timescale, they will inexorably decay. This transient regime will eventually effect the \textit{return time}, the time the system needs to return sufficiently close to the initial state. Looking at the dynamics shown in Fig. 1, one can realize that in the non-normal case{, namely in presence of topological resilience,} the return time is considerably longer than in the normal one.

\begin{figure}[h]
\centering
\includegraphics[scale=0.25]{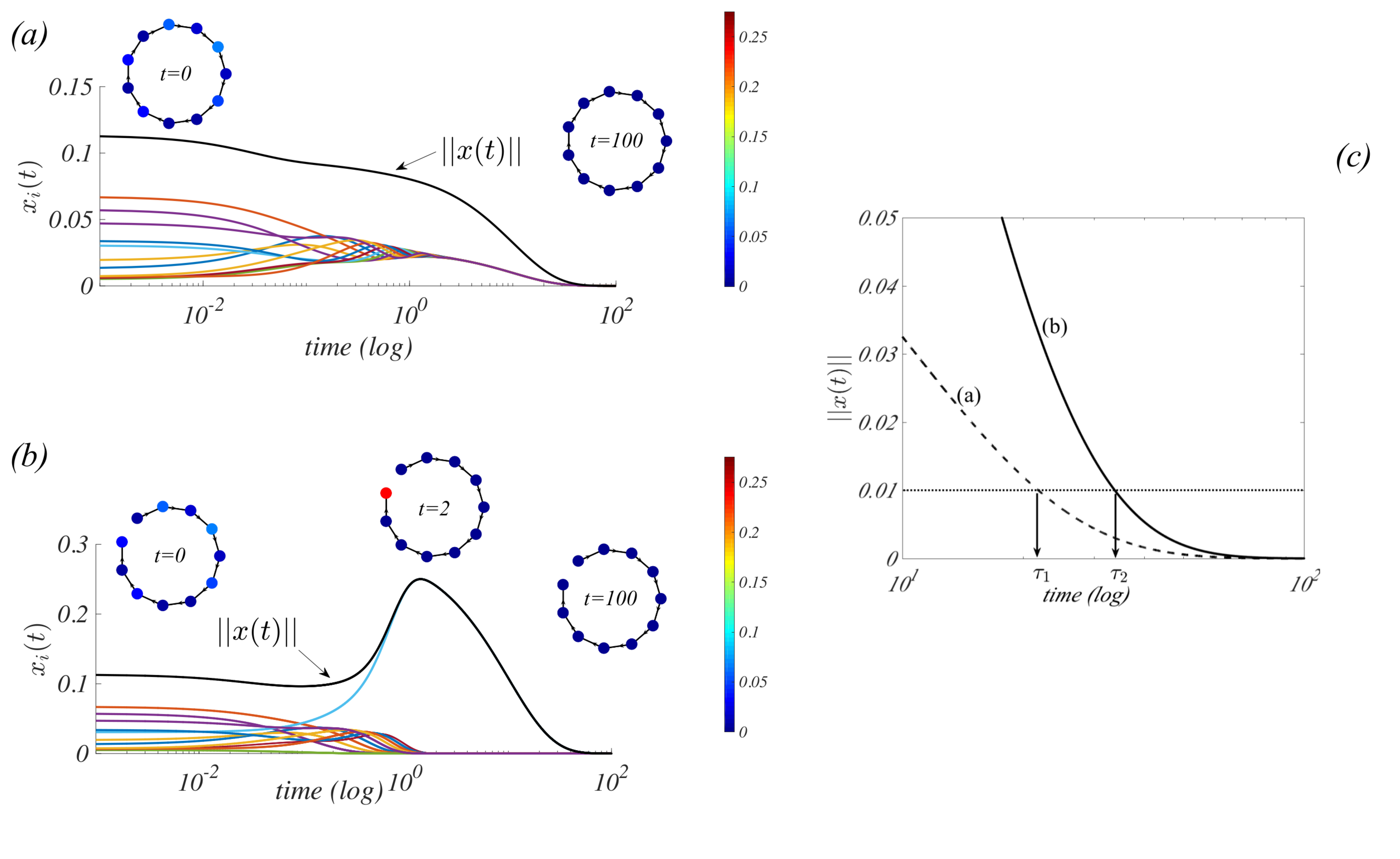}\\
\label{fig:FIG1}
\caption{\textbf{Transient growth and non-normality.} We represent the time evolution of the node density $x_i(t)$ (colored curves in panels $\textbf{(a)}$ and $\textbf{(b)}$) for the linear model, $\dot{\mathbf{x}}=-a\mathbf{x}+D\mathbf{L}\mathbf{x}$, on top of an unidirectional: $\textbf{(a)}$ ring and $\textbf{(b)}$ path, both made of $M=11$ nodes. The norm $||x(t)||$ (black curves) describes the global behavior of the system. The model in panel $\textbf{(a)}$ corresponds to a \textit{normal} system and thus, independently from the system parameters and the initial conditions close to the equilibrium point, the perturbations are doomed to die out. On the other hand, using a \textit{non-normal} network $\textbf{(b)}$, and assuming diffusion to be faster than the capture rate into the nodes sinks ($D\gg a$), then the density on the terminal node experiences a strong transient amplification before reaching definitively the equilibrium. This transient regime will influence also the \textit{return time} (see panel $\textbf{(c)}$) where the return time $\tau_2$ corresponding to the non-normal network in $\textbf{(b)}$ is larger than the return time $\tau_1$ associated to the network used in $\textbf{(a)}$. For both networked systems we used the same initial conditions
and the parameters $a=0.1$ and $D=10$.} 
\end{figure}

Different measures do exist to quantify the non-normality of {a matrix}~\cite{spectra,neubertmurray} and thus to unravel the accompanying behavior; for instance, in numerical analysis the \textit{condition number} is used to measure the accuracy of the solution of a matrix linear equation \cite{golub}, in the linear operator theory in Banach spaces the \textit{spectral abscissa} measures the growth bound \cite{kato} and in population dynamics \cite{ecoresilience} the \textit{reactivity} is used for evaluating the ecological resilience. The latter one, hereby referred to as the \textit{numerical abscissa} \cite{spectra}, suits our purpose; for a given matrix $\textbf{A}$ it is defined by:
\begin{equation}
\omega(\textbf{A}) = \mathrm{sup}\; \sigma\bigg((\textbf{A}+\textbf{A}^*)/2\bigg)\, ,
\label{eq:num_abs}
\end{equation}
where $\sigma(\textbf{A})$ denotes the spectrum of the matrix $\textbf{A}$ and $\textbf{A}^*$ is its conjugate transpose. Let assume $\textbf{A}$ to be a stable matrix \cite{golub}, if the numerical abscissa takes negative values then the orbits exponentially approach the fixed point, on the other hand if $\omega{(\textbf{A})}>0$, then a transient growth occurs whose size is proportional to the numerical abscissa, determining thus the magnitude of the non-normality of $\textbf{A}$ (see Fig. 1 and the Supplementary Material for more details).

Eventually, we will use the Euclidian norm $||\textbf{x}||=\sqrt{x_1^2+...+x_M^2}$ to have an overall evaluation of the system evolution as a single entity. Based on the above remark, we observe that for a normal matrix, the Euclidean norm of a solution of the associated linear matrix system, should always steadily decrease exponentially when the system is initialized close to a stable state. On the other hand, in a non-normal case and still using the same initial conditions, $||\textbf{x}||$ first experiences a transient amplification before decaying to zero. This temporary growth is proportional to the non-normality of the involved matrix as it can appreciated looking at the explicit solution $\mathbf{x}(t)=c_1\mathbf{v}_1+c_2t^{M-2}e^{\lambda_2 t}\mathbf{v}_2+c_3 e^{\lambda_3 t}\mathbf{v}_3$, where $c_j$ are constant set by the initial conditions, $\lambda_2$ and $\lambda_3$ the negative eigenvalues of the non-normal matrix, the former with multiplicity $M-2$, and $\mathbf{v}_j$ the associated eigenvectors ($\lambda_1=0$).

The key point is now to translate the information of non-normality onto the networks, said differently which are the networks that can manifest non-normal dynamics? A complete and exhaustive answer has not yet been provided, although appropriate models that involve the non-normality of linear operators have already been identified. Among them, we can mention the Hatano-Nelson model \cite{hatano}, the Fibonacci sequences \cite{embree}, the balanced neural networks \cite{neurons} or the random triangular matrices \cite{visw}. In principle, we can say that sparse random networks whose nodes are accommodated into directed chains-like structures, {(in our case responsible for a directional flow into the network)} which translate into triangular, or with similar asymmetric properties, adjacency matrices $\textbf{A}$ have a good degree of non-normality. Let us observe that the prototypical network model introduced in Fig. 1 can be used as backbone for the required non-normal network; starting from this remark, we have thus built a directed and weighted small-world network \cite{watts}  based on the Newman-Watts mechanism \cite{NW}. More precisely, taking a random weighted directed ring, acting as the backbone, we add long-range links with random weights considerably smaller than the ones of the ring, reflecting thus the observation that the small-world topology is widely spread in Nature~\cite{newman}: direct interaction between far away nodes, is less probable and weaker with respect to closer ones. The adjacency matrix results thus almost triangular, a structure that closely resembles the Jordan blocks of the canonical form of a non-normal matrix~\cite{golub}. A generic representative network of this family is presented in Fig. 3.

\section*{Allee effect in non-normal ecological networks}

The non-normal assumption is thus responsible for unexpected outcomes on the dynamics, constituting hence a basis for the resilience of networked systems being related to the structural property of the network. To illustrate the potentiality of this mechanism we describe a major application in ecology, the Allee effect~\cite{allebook}. 

Ecosystems are fragile systems resulting thus ideal study cases for the resilience phenomenon. There are examples of species dynamics where, in case of low densities, the growth rate is negatively correlated with the population density, meaning that they need a minimum number of individuals to reproduce and survive in their habitat. The reasons for the Allee effect are found in the loss of genetic diversity of the population for low densities \cite{courchamp}, the demographic stochasticity including the sex-ratio fluctuations which impede the reproduction \cite{demo} or the reduction of the cooperative interactions when there are few individuals \cite{allee}. The latter is of course directly related to the network of interactions and as we will see to the non-normal dynamics. A perfect example of the community pro-cooperation or facilitation \cite{allebook} is the African wild dog, \textit{Lycaon pictus}, endemic and widely spread in a large geographical area of Africa \cite{courchamp}. Yet, rapid climate changes leading to the desertification and fragmentation of the habitat from the agriculture expansion are increasing the risk of extinction. Cooperation plays a crucial role in the survivability of the species being decisive in the hunting strategy, the raising of chubs or the defense from bigger predators \cite{courchamp}. 

On the other hand, Allee effect may yield also a positive effect when it manifests in the persistence of pathogens as is the case of the measles \cite{measles}. It has been observed that for measles the critical community size, namely the number of humans needed for the disease to spread overall in a population of susceptible individuals, is between $250,000$ to $400,000$. This largely explains why vaccination programs in developing countries, involving only a fraction of the total population, have nevertheless produced an appreciable decrease in the number of infected individuals.
    
In the literature, the Allee effect is often introduced by modifying some generic model of population growth, for instance the logistic equation~\cite{murray} (see Fig. 2). It is clear that species will survive only if the initial density is beyond a certain threshold $A$, there are indeed two (stable) states: an undesired one which corresponds to extinction, $x^*_1=0$ and the desired one $x^*_2=1$ where the system maximizes its opportunities to survive taking into account the available resources \footnote{Let us note that Eq. (\ref{eq:allee}) has been written using the rescaled variable $x_i=n_i/K$, namely the ratio of the number of individuals in the $i$-th patch and the carrying capacity $K$.}.

\begin{figure}[h]
\centering
\includegraphics[scale=0.35]{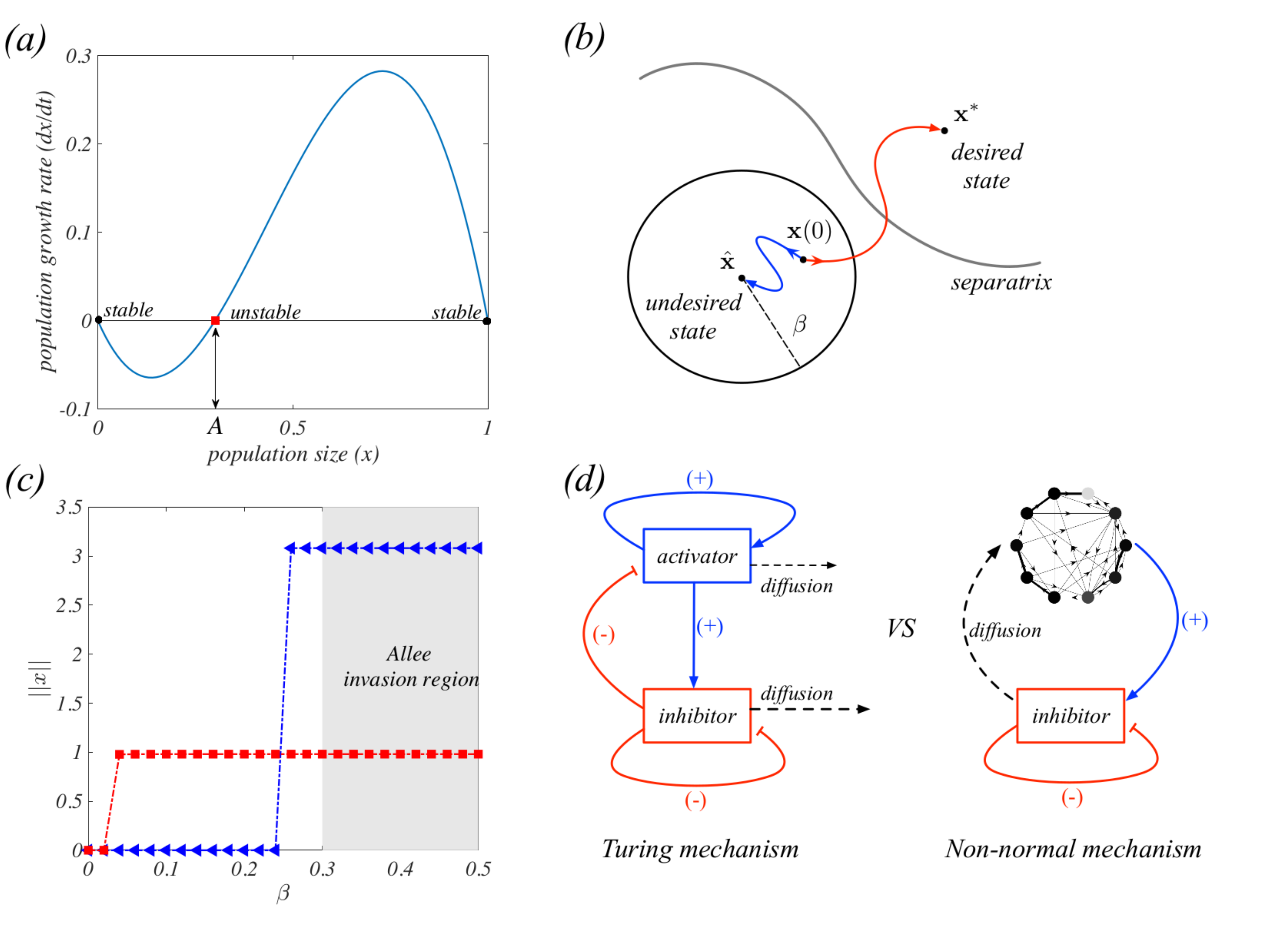}
\label{fig:FIG2}
\caption{\textbf{Resilience and the Allee effect.} $\textbf{(a)}$ The non-spatial Allee model, $dx/dt=rx(1-x)(x/A-1)$, in normalized variables, $x=n/K$ where $n$ is the population size and $K$ the carrying capacity, $r$ intrinsic growth rate and $A$ Allee critical parameter. The Allee effect states that for $x(0)<A$ the species fails to survive. $\textbf{(b)}$ Non-normal versus normal system. For the normal systems, an initial condition $\textbf{x}(0)$ converges to the stable state $\hat{\textbf{x}}$ of the attraction basin it belongs to, while it can leave its attraction basin to reach another stable solution $\textbf{x}^*$, if the transient phase is strong enough. $\textbf{(c)}$ Bifurcation diagram of the networked Allee model. For a normal system (blue), the size of the survival zone, $\sim 0.25$, is very close to that of the non-spatial model, $0.3$, on the other hand for non-normal systems (red) the survival zone is strongly increased, $\sim 0.05$. $\textbf{(d)}$ Turing pattern mechanism versus the non-normal one. In the former two species are needed, with the inhibitor diffusing faster than the activator; in the non-normal scheme only one inhibitor species is enough to create patterns once it diffuses on a non-normal network playing the role of the activator.}
\end{figure}

To go one step further in the modeling, let us assume the population to live in a patchy environment where animals can move across the niches; the dynamics can thus be described by the following diffusively coupled equations:
\begin{equation}
\frac{dx_i}{dt} = rx_i\left(1-x_i\right)\left(\frac{x_i}{A}-1\right) + D\sum_{j=1}^M L_{ij}x_j,\forall i\, ,
\label{eq:allee}
\end{equation}
where $x_i$ denotes the species density in the $i$-th patch, $r$ the reproduction rate, $A$ the Allee coefficient and $D$ the diffusion coefficient, all assumed for simplicity to be the same for all patches.

If the network of interactions is normal and the initial conditions do not exceed the Allee threshold, $x_i(0) < A, \forall i$, then the species goes extinct and diffusion cannot prevent it. Conversely if the underlying network belongs to the family presented above (see Fig. 3), the system fate turns upside down and the population will survive {reducing thus the system resilience}. The explanation for this behavior can be found in the competition between the diffusion mechanism and the reproduction rate. It can happen that the initial transient amplification induced by the non-normality is strong enough to surpass the Allee threshold, at least in some of the patches, and consequently the system saturates avoiding the extinction.
A schematic illustration of the resilience response of the Allee model on a non-normal network is presented in Fig. 2. The non-normal spatial support, will facilitate the orbits to cross the separatrix among the attraction basins, to end up in the survival state ${\textbf{x}}^*$, even if they were initially in the basin of the attraction of extinction state $\hat{\textbf{x}}=0$.

The effect of the non-normality on the resilience as discussed above, emphasizes the great advantage that the network topology represents in the population dynamics. The theoretical approach hereby discussed can be used to describe different scenarios in ecology like the introduction of new individuals in a particular habitat for the goal of species conservation or biological control from invasive species \cite{courchamp}. Often, the achievement of such goals is strongly conditioned by the Allee effect, the new introduced individuals are not quite numerous and to prevent their extinction repeated releases are necessary before the establishment of the new introduced species. However, if the dispersion locations in the ecosystem under study are chosen to induce a non-normal behavior, as previously shown, this will increase the probability of the newly introduced species to survive. 

\begin{figure}
\centering
\includegraphics[scale=0.3]{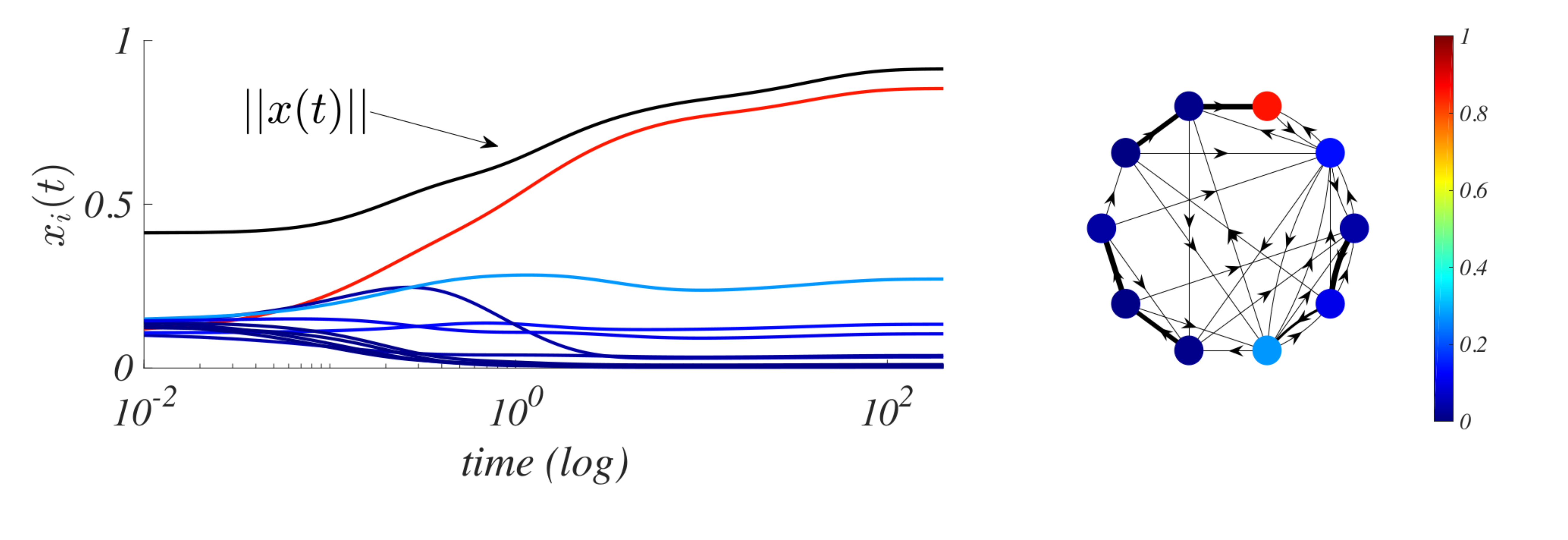}
\label{fig:FIG3}
\caption{\textbf{Patterns emergence in non-normal networked systems.} We represent the time evolution of the node density $x_i(t)$ (colored curves on the left panel) for the Allee model~(\ref{eq:allee}) on top of a non-normal network made of $M=10$ nodes (right panel), together with the norm $||x(t)||$ (black curve) to appreciate the global behavior of the system. One can observe the onset of a stable spatial pattern instigated by a ``terminal node'' (red node on the right panel and red curve on the left panel) of a backbone directed weighted path. Parameters are $a=0.1$, $A=0.3$ and $D=10$. The core ring has stronger links compared to the long-range ones {(See Supplementary Material).}}
\end{figure}

\section*{A new mechanism for pattern formation}

The previously discussed model where the non-normality facilitates the survival of the species and avoid the Allee effect does in fact, yield a non homogeneous non-linear pattern. One of the mostly diffused mechanism responsible for the pattern formation, is the one introduced by A. Turing in his seminal paper on morphogenesis \cite{turing}. He stated that a non-linear reaction-diffusion system, starting from a homogeneous stable state, can experience a diffusion-driven instability producing spatially inhomogeneous solutions, the celebrated Turing patterns \cite{turing,murray}. For this mechanism to be able to realize Turing patterns, the minimal requisites are the presence of at least two species, one being ``activator'' (capable to trigger their own growth) and the other ``inhibitor'' (antagonist to the former, impede any further growth once diffusing), and moreover the ratio of their diffusion coefficient (inhibitor vs. activator) should be larger than some threshold, which in realistic models is of the order of $\sim 10$ \cite{murray} (see Fig. 3). 

Even though the advantage of the directed network over undirected ones, have been emphasized in the process of formation of Turing patterns~\cite{asllani}, the non-normality of the spatial support has not yet been considered. In this new scenario, this topological feature of the network of interactions will force the inhibitors in some of the nodes, to initially increase their concentration until they saturate in the non-linear phase. What is remarkable is that the species which apparently tends to go extinct because of the negative growth rate, exploit a faster diffusion process that makes the species to spread before the individuals counteract, and eventually lead to self-organization. This is thus a new mechanism, different form the Turing one, capable to explain the pattern formation process, observe moreover that one species is enough to activate this phenomenon, provided it diffuses on a non-normal network.

\section*{Discussion}

We studied the role of non-normal topology in the resilience phenomenon of dynamical systems defined on directed networks. Interesting enough the way parts of interconnected systems interact could make them vulnerable to weak external perturbations. In the case of non-normal networks, unexpectedly, these perturbations will follow an initial amplification that can lead the system to a new state, possibly far from the initial one. To increase the robustness of the networked system, for instance in the case of power grids once we want to reduce cascade failures, one should reduce the non-normality.

Once the non-normality guides the system toward a non-homogeneous equilibrium state, the process at work can be cast in the framework of patterns formation driven by dynamical instability. The single species case hereby presented shows the impact of the network topology on the self-organization process, allowing the formation of patterns beyond the Turing mechanism.
 
This approach may thus shed light onto problems related to multi-species models in ecology; for instance, if the goal is to promote coexistence or survival of newly introduced species, then the network of interactions should be designed to be as much as possible non-normal to prevent thus the Allee effect to favor the extinction. On the other hand in the case the invasive species is an infectious pathogens~\cite{invasive} (noxious Allee effect), then one should avoid or reduce non-normality.
 
This last remark can thus have a very relevant impact on the design of urban agglomerations and their transportation networks, in order to increase their robustness against the spreading of diseases. As a particular study case we mention (see the Supplementary Material) the London Tube transport system, where the daily commuting fluxes {from the outskirts to the city center,} create a non-normal system where the possibility for pathogens to survive and the consequent epidemics outbreak, are highly enhanced compared to people living and moving in the downtown.

Let us observe that the non-normal mechanism hereby studied is way more general than the few applications presented, as one can infer from system Eq.~(\ref{eq:general}). In particular, we decided to focus onto the diffusion process and the role of space; nevertheless one can provide interesting applications also to non-spatial systems where the non-normality shapes the interactions among individuals of different species, for instance in the framework of foodwebs (see Table 1).

{In conclusion we are confident that the non-normality property can determine multiple and still unexplored achievements in the dynamics of networked system.}

\section*{Acknowledgments}
The work of M.A. and T.C. presents research results of the Belgian Network DYSCO, funded by the Interuniversity Attraction
Poles Programme, initiated by the Belgian State, Science Policy Office. The work of M.A. is also supported by a FRS-FNRS
Postdoctoral Fellowship. The authors acknowledge also D. Fanelli and G. Deffuant for reading the manuscript.
%
%
%

%
%
%

\appendix
\section*{Supplementary material}

\section*{Materials and Methods}

\vspace*{0.5cm}

\subsection*{Non-normality measures and transient growth}

Non-normal linear dynamical systems exhibit a peculiar time evolution, the transient phase can be very different from the case of normal ones; the aim of this section is to introduce and discuss some basics facts and measures related to non-normal systems, the interested reader can consult [8] for a more detailed discussion. We hereby limit ourselves to the case of continuous time dynamical systems (Ordinary Differential Equation), the reader must however be aware that an analogous theory exists for discrete time dynamical systems (maps). Let us thus start by considering a $M\times M$ real matrix $\textbf{A}$ with a non-positive spectrum and investigate its associated linear ODE, $\frac{d\textbf{x}}{dt}=\textbf{Ax}$, whose solution is given by $\textbf{x}(t)=e^{t\textbf{A}}\textbf{x}(0)$, being $\textbf{x}(0)$ the initial condition. In order to study the behavior of the latter system we can look at the time evolution of the Euclidian norm of the solution $||\textbf{x}|| = \sqrt{x_1^2+x_2^2+\dots+x_M^2}$.

Let us start by analyzing the short time behavior of the solution, namely to study the limit $t\rightarrow 0^+$. To this aim one can define the \textit{numerical abscissa}, $\omega(\textbf{A})$, also named in the ecology literature as \textit{reactivity} [11]: 
\begin{equation}
\omega(\textbf{A}):=\max_{||\textbf{x}_0||\neq 0}\left[\left(\frac{1}{||\textbf{x}||} \frac{d ||\textbf{x}||}{dt}\right)\bigg |_{t=0}\right]\, ,
\label{eq:num_absc} 
\end{equation}
where for a sake of notation we didn't emphasize the dependence of the solution $\textbf{x}$ on the initial condition $\textbf{x}_0$.

The value of the numerical abscissa can be calculated as follows:
\begin{eqnarray}
\max_{||\textbf{x}_0||\neq 0}\left[\left(\frac{1}{||\textbf{x}||} \frac{d ||\textbf{x}||}{dt}\right)\bigg |_{t=0}\right]&=&\max_{||\textbf{x}_0||\neq 0}\left[\left(\frac{1}{||\textbf{x}||} \frac{d\sqrt{\textbf{x}^*\textbf{x}}}{dt}\right)\bigg |_{t=0}\right]\nonumber\\
&=&\max_{||\textbf{x}_0||\neq 0}\left[\left(\frac{\textbf{x}^* d\textbf{x}/dt + \left(d\textbf{x}/dt\right)^*\textbf{x}}{2||\textbf{x}||^2}\right)\bigg |_{t=0}\right]\nonumber\\
&=&\max_{||\textbf{x}_0||\neq 0}\left[\left(\frac{\textbf{x}^* \left(\textbf{A} + \textbf{A}^*\right)\textbf{x}}{2||\textbf{x}||^2}\right)\bigg |_{t=0}\right]\nonumber\\
&=&\max_{||\textbf{x}_0||\neq 0}\left[\frac{\textbf{x}_0^* H(A) \textbf{x}_0}{\textbf{x}_0^*\textbf{x}_0}\right]\, ,
\label{eq:react}
\end{eqnarray}
where $\textbf{A}^*$ is the conjugate transpose of the matrix $\textbf{A}$ and we have denoted the Hermitian part of $\textbf{A}$, by $H(\textbf{A})=\left(\textbf{A} + \textbf{A}^*\right)/2$. According to the Rayleigh's principle [20] the right hand side of Eq.~(\ref{eq:react}) is equal to the largest eigenvalue of the $H(\textbf{A})$, hence we can conclude that $\omega(\textbf{A})=\sup\sigma(H(\textbf{A}))$. Notice that this measure does not depend anymore on the initial conditions, meaning that it characterizes the intrinsic properties of the matrix $\textbf{A}$ and not of a specific solution. 

The second limit we are interested in, is the long time behavior, namely $t\rightarrow +\infty$. This can be studied using the \textit{spectral abscissa} of the matrix $\textbf{A}$, hereby denoted by $\alpha(\textbf{A})$ and defined by:
\begin{equation}
\alpha(\textbf{A}):=\max_{||\textbf{x}_0||\neq 0}\left[\lim_{t\rightarrow +\infty}\left(\frac{1}{||\textbf{x}||} \frac{d ||\textbf{x}||}{dt}\right)\right]\, .
\label{eq:spec_abs} 
\end{equation}
It is well known that the eigenvalue with the largest real part completely determines the asymptotic behavior of the solution of the ODE, thus one can compute the {spectral abscissa} as $\alpha(\textbf{A})=\sup \Re(\sigma (\textbf{A}))$.

We can thus conclude this part by observing that, although $\alpha(\textbf{A})<0$, if $\omega(\textbf{A})>0$ then the equilibrium will be stable but a transient growth will emerge in the short time regime producing a deviation from the steady exponential decay of stable normal systems.

In the end we analyze, for illustrative purpose, the time evolution of the norm of the solution of the linear ODE system involving the following matrices $\textbf{A}_i$, $i=1,2$
\begin{equation*}
\textbf{A}_1=\begin{pmatrix}
-1 & 1\\0 & -2
\end{pmatrix} \;\;\;\; \mathrm{and}\;\;\;\;
\textbf{A}_2=\begin{pmatrix}
-1 & 10\\0 & -2
\end{pmatrix}.
\end{equation*}
A straightforward computation allows to obtain the numerical abscissa, $\omega(\textbf{A}_1)=-0.79<0$ and $\omega(\textbf{A}_2)=3.52>0$, and the spectral abscissa $\alpha(\textbf{A}_1)=\alpha(\textbf{A}_2)=-1$. Although the latter is the same for both matrices, the existence of a positive numerical abscissa for the second matrix determines a transient growth to occur (see main panel Fig. 4). A schematic illustration of the impact of these two measures on the transient time evolution of a solution is given in the inset of Fig. 4. Once again, this example demonstrates that the spectrum alone is not able to describe the linear dynamics in the setting of non-normal dynamical systems. 

\subsection*{A model to generate non-normal networks}

We have shown that dynamical systems running on top of networks whose adjacency matrix (or Laplacian matrix) satisfies a non-normal condition, can exhibit unexpected dynamical behaviors. We hereby present a general model to construct such non-normal networks, however before to proceed with the details, let us illustrate the intuition behind the construction.

As already pointed out in the main text, in order to have a transient growth dynamics, it is necessary that the matrix of the associated linear dynamical system, $\dot{\textbf{x}}=\textbf{Ax}$, should be non-normal (and stable); namely the matrix $\textbf{A}$ does not own a proper basis of eigenvectors and as consequence there exists at least one eigenvalue, $\lambda^*$, with a geometric multiplicity higher than one, $m>1$, which implies an initial transient growth of the kind $t^{m-1}e^{t\lambda^*}$. This example brings to the most simple non-normal network where the adjacency matrix $\textbf{A}$  corresponds to a $m$-dimensional Jordan block resulting thus in a $1D$ unidirectional path, that can be seen as an unidirectional "broken" ring (see panel b Fig. 2).

Despite its simplicity, this network can be used as basic building block for a more general model of non-normal networks, roughly inspired to the Newman-Watts small-world model [25]. We start thus with a unidirectional weighted $1D$ ring where weights are chosen from the uniform distribution $U[0,\gamma]$ with $\gamma>1$, that will play the role of directionality parameter; let us assume to order nodes such that the existing links connect node $i$-th with node $(i+1)$-th. Let us observe that the ring is closed, however due to the different weights, some weak link could act as an effective break, enhancing thus the non-normality. This core directional network mimics the fact that agents are all forced to move in the same direction. However, for a sake of generality, one may consider agents to still have some inertia to move in the opposite direction, even if with a very low probability; to reproduce this fact, we assume that weak links, whose weights are of order $1$, may also exist pointing in the opposite direction. More precisely, for all $i=1,\dots,N$ with probability $p_1<1$ we create a weak link from node $(i+1)$-th pointing to node $i$-th. Finally, in complete analogy with the small-world model, long-range links do exist with weights still of order $1$; more precisely for all couples of node $(i,j)$ such that $|i-j|\geq 2$, we add with probability $p_2<1$ a directed link.

The resulting adjacency matrix, although, not being exactly a Jordan block, will result strongly non-normal, because of the large values of $\gamma$ and of the wide distribution of the weights on the main ring with respect to the remaining ones. The algorithm can also be easily extended to a structure where multiple Jordan blocks are present or by defining the weights of the long-range links to be inversely proportional to the {geodesic} distance. Note that the adjacency matrices generated so far by the algorithm are non-normal, but not stable in general. In order to observe the transient growth in the short time dynamics, it is necessary to assume also the matrix to be stable. This latter fact is surely achieved once one considers the full dynamical system, namely taking into account also the reaction part and the Laplacian matrix.

In Fig. 5 we show three different examples of non-normal networks made by $M=10$ nodes with $p_1=0.5$ and $p_2=0.2$ for three different values of the parameter $\gamma=4$ (weakly non-normal), $\gamma=10$ (non-normal) and $\gamma=16$ (strongly non-normal). As one can appreciate as $\gamma$ increases, the numerical abscissa $\omega(\mathbf{A}(\gamma))$ also gets, on average, larger. We decided to let $\gamma$ to vary because its impact is stronger than the one of $p_1$ (as $p_1$ increases the Jordan blocks tend to have smaller size) and $p_2$ (as $p_2$ increases the matrix becomes denser).

\vspace*{0.5cm}

\section*{Supplementary Text}

\vspace*{0.5cm}

\subsection*{Outbreak of epidemics in the commuters transport networks. A toy model based on the London Tube}

We hereby consider a second relevant application of non-normality in the framework of epidemics spreading on metropolitan scales. We will analyze an abstract spreading model based on the London Tube network and we will emphasize the role of commuters during the peak hours coming from the outskirts of Greater London, in the outbreak of the epidemics of some pathogens that manifest a strong Allee effect, such as the measles virus.

We will consider two principal lines of the London Tube (Fig. 6) that share different topologies and transport features. The first one is the yellow Circle Line (bottom right panel Fig. 6) which encircles the city downtown and has intersections with almost all the other lines. It is geographically dense, namely the stations are very close each other, and it is one of the lines of London underground system with the largest number of transported passengers, making it a perfect example to test epidemics spreading.
The hypothetical pathogen we are considering, say the measles, is strongly affected by the Allee effect, so in order for the epidemics to outbreak, we need a critical number of vectors, namely infected humans, in the same place for a given amount of time. In Fig. 7 panel (a), we show the result of the simulation of the spatial Allee model once the underlying network is represented by the yellow Circle Line (Eq. (3) in the main text); because only people traveling in the same train can get infected, we consider only one direction of the line and thus assume the network to be directed. Moreover, we assume that, on average, the number of people in the train to remain constant for the time interval we consider, namely because of the density of the stations and of the centrality of the line, roughly the same number of person gets in and gets off the train at each station. The random variations from the average is responsible for the the fluctuations visible in the Fig. 7. The main outcome one can appreciate from the panel (a) of Fig. 7 is that the disease does not persist and the epidemics will not outbreak, at odd with the intuition that the circle line will support many commuters and thus inducing an high chance for the virus to survive due to the large number of human encounters.

On the other hand, if we consider the magenta Metropolitan Line (bottom left panel of Fig. 6) which connects the northwestern outskirts with the downtown of London, and we perform a similar hypothetical experiment of virus spreading, the result overturns (see Fig. 7 panel (b)). Here we assume that passengers for the period of time we considered, mostly do not get off the train until they reach the center of the city. This induces a strong directionality in the network, which will result non-normal, and thus allowing the virus to overcome the Allee threshold. It is now clear from the simulation that the epidemics outbreaks; once the commuters reach their destinations at the center of the city they are almost all infected and then they will contribute to spreading the virus in the central part of the city and then, as a consequence, in the whole city.

This phenomenon can be easily explained in the framework of non-normal dynamics. The Circle Line can be seen as a normal network and moreover commuters spend, on average, a short time on these trains, independently from the fact that they are vectors of the virus or not; in conclusion the pathogen does not has sufficiently time to develop and reach numbers allowing to invade the population where it diffuses. On the contrary, possible vectors (commuters) living in the suburbs of the city and spending a longer time in the densely populated train, make possible the spread of the virus, because of the non-normality structure of the network, as in the case of the Metropolitan Line.

In conclusion, with this toy model we have been able to show that independently from the topology of the spatial domain, if - by their actions - the individuals determine a strongly directed flux from one region to another, then the spreading pathogen will avoid the Allee effect and so helping its survivability.

\subsection*{Details of the numerical simulations}
\underline{Figure 1}\\
The results reported in the panel of such Figure have been obtained using a deterministic numerical integration scheme based on a $4$-th order Runge-Kutta method. For both panels a) and b) the initial conditions were chosen randomly from a uniform distribution $[0, 0.1]$. 
For both networks, the ring and the path, the links do all have weights $1$.
The return time has been defined as the time needed by the system to retour $0.01$-close to the initial condition.\\
\underline{Figure 2}\\
To obtain the bifurcation diagram shown in panel c) we considered initial conditions close to the ones used for Fig. 1, that is drawn from a uniform distribution $U[0, 0.1]$, then we numerically simulated the spatial Allee model on top of a non-normal network (red squares) and of a weakly non-normal one (blue triangles) and eventually we determined the asymptotic size of the system described by $||\mathbf{x}||$. The model parameters have been set to $r=0.1$, $A=0.3$ and $D=10$.
More precisely for several values of $\beta\in[0,0.5]$ we draw initial conditions from the uniform distribution $U[0, 0.1]$, and then we add a random number drawn from the distribution $U[\beta-0.05,\beta+0.05]$; to reduce the variability in the results induced by the randomness of the initial conditions, we repeat $10$ times the same experiment with the same $\beta$.   
The non-normal network has been built using the method presented above with parameters $M=10$ nodes, $\gamma=10$, $p_1=0.5$ and $p_2=0.2$, resulting into a numerical abscissa $\omega(\textbf{J})=1.87$. The weakly non-normal network has been obtained similarly but using $p_2=0.8$, giving a numerical abscissa $\omega(\textbf{J})=0.57$. Let us observe that we hereby compute the numerical abscissa using the Jacobian of the dynamical system and not only of the network adjacency matrix.\\
\underline{Figure 3}\\
The results reported in this Figure have been obtained using a deterministic numerical integration scheme based on a $4-$th order Runge-Kutta method. The initial conditions were chosen randomly from a uniform distribution $[0, 0.1]$. The model parameters have been set to $a=0.1$, $r=0.1$, $A=0.3$ and $D=10$.
The non-normal network has been built using the method presented above with parameters $M=10$ nodes, $\gamma=10$, $p_1=0.5$ and $p_2=0.2$.

\clearpage

\section*{Supplementary Table}

\begin{table}[h]
\begin{tabular}{l |c|c|c|c|c}
  \hline
Name & N of nodes & N of links & Non-normal & $\omega(\mathbf{A})$ & Ref. \\
  \hline
  Chesapeake (Summer carbon flows) & $36$ & $177$ & yes & $4.11\, 10^5$ & \cite{BairdUlanowicz1989}\\
  Lower Chesapeake (Summer carbon flows) & $34$ & $178$ & yes & $1.06\, 10^5$ & \cite{Hagy2002}\\
  Middle Chesapeake (Summer carbon flows) & $34$ & $209$ & yes & $1.65\, 10^5$ & \cite{Hagy2002}\\
  Upper Chesapeake (Summer carbon flows)  & $34$ & $215$ & yes & $6.23\, 10^4$ & \cite{Hagy2002}\\
  Crystal River (1) & $21$ & $125$ & yes & $2.52\, 10^3$ & \cite{Ulanowicz1986}\\
  Crystal River (2) & $21$ & $100$ & yes & $1.94\, 10^3$ & \cite{Ulanowicz1986}\\
  Everglades Graminoid Marshes & $66$ & $916$ & yes & $1.44\, 10^3$ & \cite{UlanowiczEtAl2000}\\
  Florida Bay Trophic Exchange Matrix & $125$ & $2016$ & yes & $2.18\, 10^2$ & \cite{UlanowiczEtAl1998}\\
  Charca de Maspalomas, Gran Canaria & $21$ & $82$ & yes & $5.40\, 10^5$ & \cite{AlmuniaaEtAl1999}\\
  Narragansett Bay Model & $32$ & $220$ & yes & $8.06\, 10^5$ & \cite{MonacoUlanowicz1997}\\
  St Marks River (Florida) Estuary & $51$ & $356$ & yes & $1.40\, 10^2$ & \cite{BairdEtAl1998}\\
  \hline  
\end{tabular}
\caption{\textbf{Food webs}. We report some figures for a set of already studied food webs. They all result to be (weighted) non-normal networks, namely $\mathbf{A}\mathbf{A}^*\neq \mathbf{A}^*\mathbf{A}$, and they posses a positive numerical abscissa. }
\end{table}

\clearpage

\section*{Supplementary Figures}

\begin{figure}[h]
\centering
\includegraphics[width=0.7\textwidth]{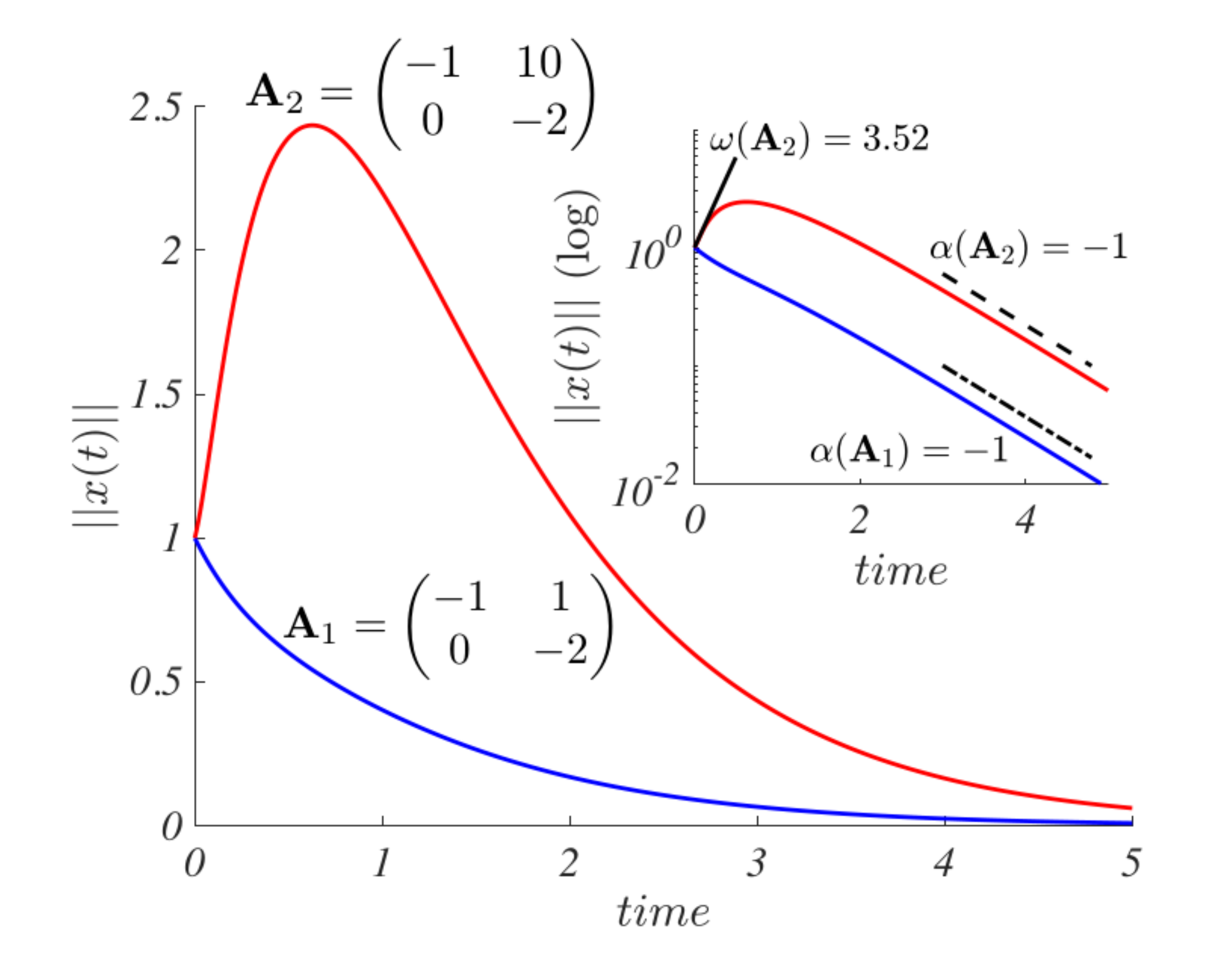}
\caption{\textbf{Time evolution of the norm of the solution of the linear ODE $\dot{\textbf{x}}=\mathbf{Ax}$}. The red curve corresponds to a non-normal matrix $\omega(\mathbf{A}_2)=3.52$ while the blue curve to a normal one $\omega(\mathbf{A}_1)=-0.79$, one can clearly appreciate the temporal growth arising in the former case. In the inset we still report the norm of the solution but in logarithmic scale to emphasize the short time behavior described by the numerical abscissa (the straight black line has slope $3.52$) and the long time one related to the spectral abscissa (the dashed and dot dashed straight lines have slope $-1$).}
\label{fig:Totsupp_fig1}
\end{figure}

\begin{figure}[h]
\centering
\includegraphics[width=0.7\textwidth]{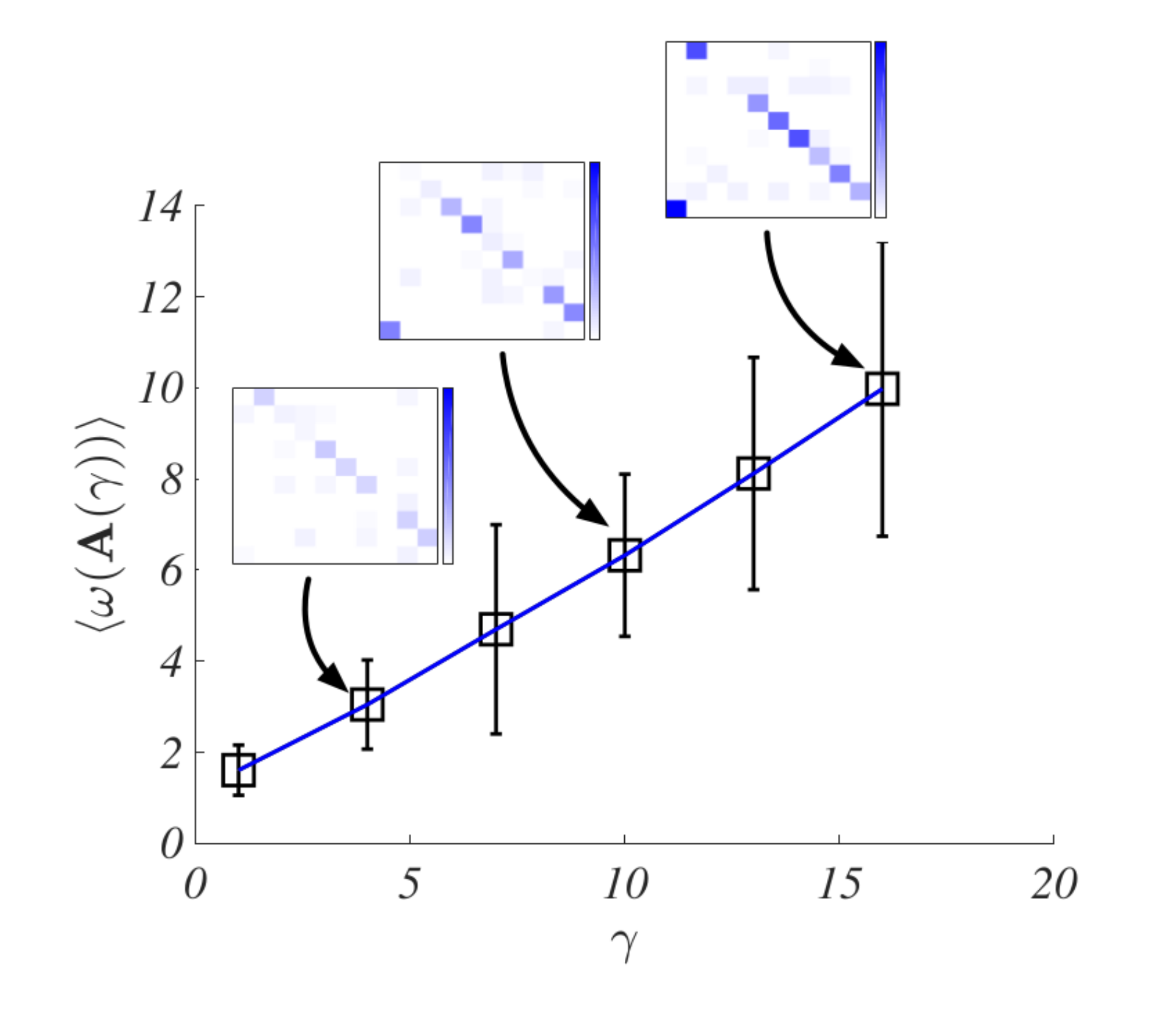}
\caption{\textbf{Numerical abscissa as a function of the directional parameter $\gamma$}. Each point has been obtained using $100$ replicas using the same set of parameters and the associated error bars are reported. For $\gamma=4$, $\gamma=10$ and $\gamma=16$ we represent three representative adjacency matrices, the darker the blue the larger the value of the weight. One can appreciate the presence of a ring-like structure: the weights on the upper diagonal are quite strong as well as the weight connecting the nodes $N$ and $1$. (bottom left corner in the matrices).}
\label{fig:nonnormaldiag}
\end{figure}

\begin{figure}[h]
\centering
\includegraphics[width=0.9\textwidth]{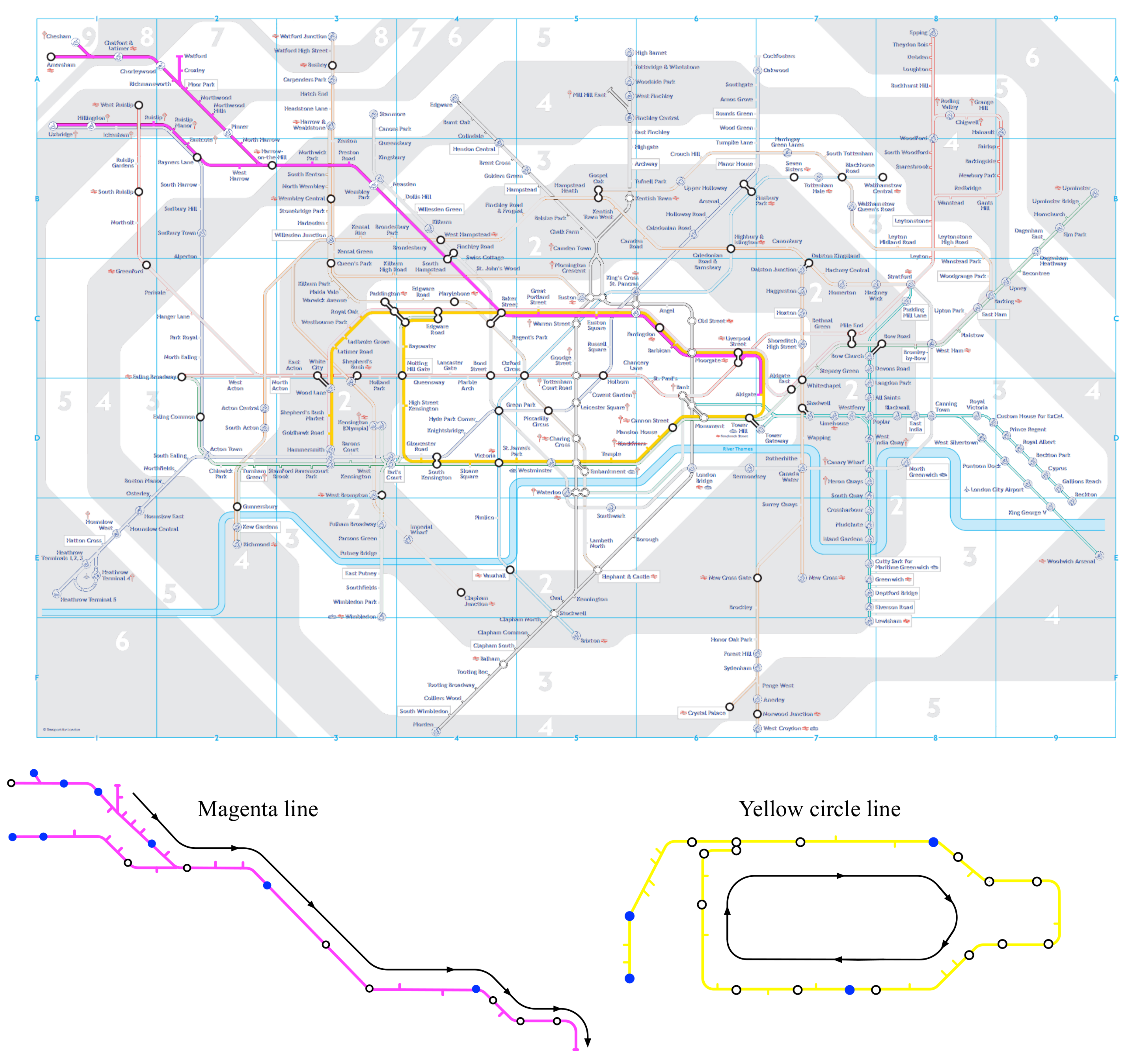}
\caption{\textbf{The London Tube map}. Top panel: the whole London tube map where the yellow Circle Line and the magenta Metropolitan Line have been emphasized to show their different geographical emplacement and shape. In the bottom panels we separately present the Magenta line (left) and the Yellow line (right) and their respective flows of commuters.}
\end{figure}

\begin{figure}[h]
\begin{center}
\subfloat[]{
\includegraphics[width=0.5\textwidth]{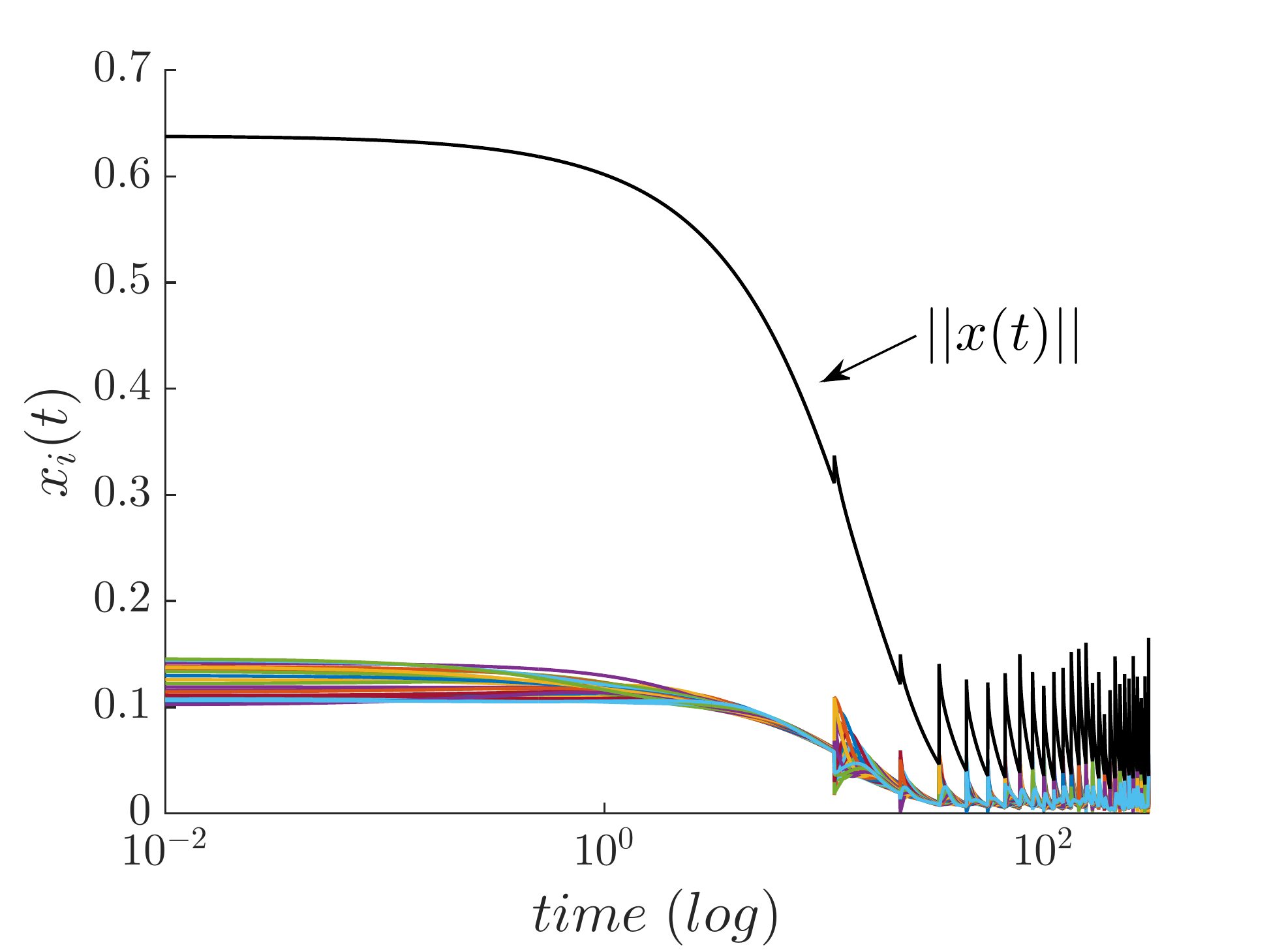}}
\subfloat[]{
\includegraphics[width=0.5\textwidth]{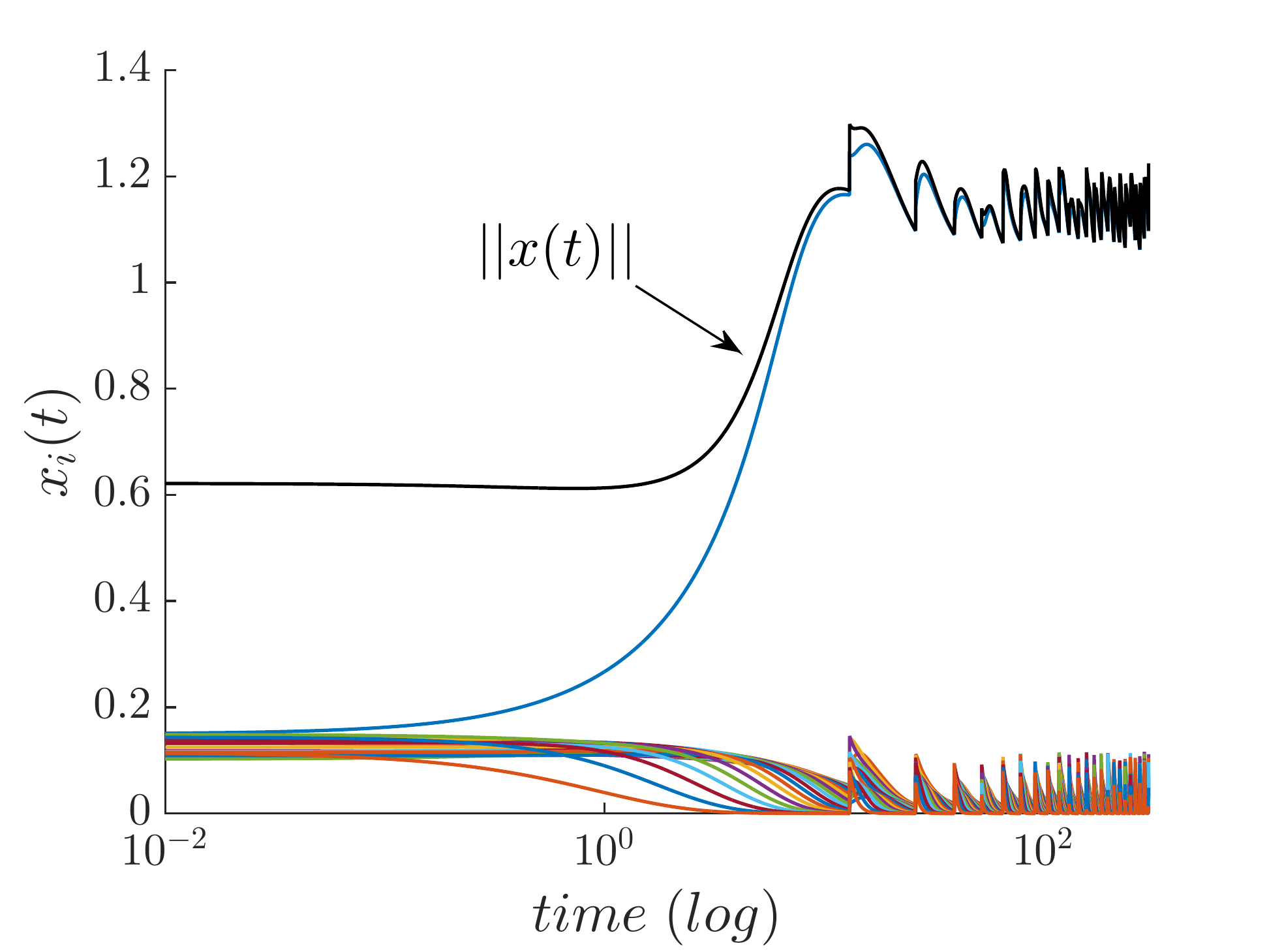}}
\end{center}
\caption{\textbf{{Hypothetical} Measles outbreak in the London Tube network.} Time evolution of the (normalized) number of infected individuals during the peak hours in the yellow Circle Line {taking into account $27$ stations} (panel $\textbf{(a)}$) and in the magenta Metropolitan Line (panel $\textbf{(b)}$) {considering $23$ stations with the fluxes} illustrated in Fig. 6. In the former case (panel $\textbf{(a)}$) we assumed the average number passengers to constant because in each station, the average number of passengers getting on equals that of those getting off. The circular topology and the Allee effect impede the outbreak of the epidemics. On the contrary, for the second case (panel $\textbf{(b)}$), we assume that, on average, as long the trains approach the downtown, the number of passengers increases, because most of the individuals have their destination in the city center. In this case, despite the presence of the Allee effect, the topology contributes to the outbreak of the measles epidemics once the train reaches the center. The parameters of the Allee model are $a=1$, $A=0.3$ and $D=10$ for both cases.}
\end{figure}

\clearpage

\bibliography{Non-normal.bib}
\bibliographystyle{Science}
\end{document}